\documentclass{mem}
\usepackage{natbib}\usepackage{txfonts}\usepackage{balance}
\usepackage{flushend}
\usepackage{graphicx}
\usepackage[breaklinks,pdftex,colorlinks=true, citecolor=blue]{hyperref}
\usepackage{xcolor}
\idline{94}{88}

\begin{document}

\title{Pulsating stars in Local Group dwarf galaxies}
%\subtitle{}
\author{C. E. \,Mart{\'{i}}nez-V{\'a}zquez\inst{1}}
\institute{Gemini Observatory/NSF's NOIRLab, 670 N. A'ohoku Pl., 96720 Hilo, HI, USA
\email{clara.martinez@noirlab.edu}
}

\authorrunning{Mart{\'{i}}nez-V{\'a}zquez}

\titlerunning{Pulsating stars in Local Group galaxies}

\date{Received: 04 October 2023; Accepted: 21 November 2023}

\abstract{

The popularity of pulsating stars resides in their capacity of determining several crucial and relevant parameters such as heliocentric distances, ages, metallicity gradients and reddening. RR Lyrae stars are old stellar tracers and have been detected in nearly all nearby galaxies that have been searched for these stars, with just a few exceptions of very low mass dwarfs. Less common but also of great importance are Anomalous Cepheids, indicators of either old or intermediate-age population, depending on their stellar origin. Classical Cepheids are only found within young stellar populations, and because of their brighter absolute magnitudes, they can be detected in galaxies farther than the Local Group.
This paper presents a concise review built upon the aforementioned pulsating stars in Local Group dwarf galaxies and some of their applications to infer important properties of their host galaxies.

}
\maketitle{}

\section{Introduction}
Pulsating stars are considered standard candles because they obey well established period-luminosity relations in optical and infrared bands \citep{CatelanSmith2015}. Therefore, they are used very often to derive accurate and precise distances. Coming from very different population, Cepheids (Population~I) and RR Lyrae stars (RRLs; Population~II) can be used as one of the first rungs in the cosmic distance ladder thanks to their recurrent and periodic brightness variation.

By far, the most frequent and common pulsating stars found among Local Group (LG) dwarf galaxies are RRLs \citep[][and references therein]{MonelliFiorentino2022}. The discovery of the first RRL star was made by Williamina Fleming  in the late 19th century through her analysis of plates from Solon Bailey’s cluster survey in 1893 \citep{Pickering1901}. Thanks to their well-know period-luminosity relation (also known as Henrietta Leavitt's Law), the popularity of this type of stellar pulsator has increased exponentially, especially in recent decades. The extensive detection of RRL stars in surveys like ASAS, Catalina, DES, Gaia, OGLE, PanSTARRS or ZTF, coupled with their role as stellar tracers of ancient stellar population \citep{Walker1989,Savino2020}, positions them as powerful tools to study old stellar structures. In addition, RRLs have been frequently used to detect/confirm new low mass (ultra-faint) and low surface brightness (ultra-diffuse) dwarf galaxies and to obtain accurate distances to those systems, where the large contamination by field stars makes challenging the determination of distances from isochrone fitting \citep[see e.g.,][]{Bechtol2015,DrlicaWagner2015,Torrealba2016,Torrealba2019}. In addition, in those systems where the rate of RRLs is statistically significant, RRLs can  be utilized as metallicity tracers, giving insight into the chemical evolution of the old population they belong to (see e.g., Sculptor - \citealt{MartinezVazquez2015,MartinezVazquez2016a}; Eridanus~II - \citealt{MartinezVazquez2021b}). Thus helping us not only to reveal the formation history and chemical evolution of their host galaxy but also to provide clues about the contribution of dwarf galaxies in the formation of the halo of the larger galaxies they are bound to.

Among the brightest and most easily recognized pulsating stars in LG galaxies are Classical Cepheids (CCs) and Anomalous Cepheids (ACs). However, these two types of pulsators belong to different stellar populations. While CCs are young ($\sim100-300$~Myr, \citealt{DeSomma2020}), ACs can belong to either old ($\gtrsim 10$~Gyr) or intermediate-age ($\sim1-6$~Gyr) population since ACs can form through two different channels: originating either from the evolution of blue straggler stars (binary channel), or else from the evolution of single, metal-poor (Z $<$ 0.0004, \citealt{Fiorentino2006}), relatively young ($\lesssim$ 6~Gyr) stars, respectively. Interestingly, \citep{Mateo1995} reported a correlation between the frequency of ACs (known as \textit{log S}) and the absolute $V$ magnitude of the host galaxy (M$_V$), while \cite{FiorentinoMonelli2012} found that the frequency of ACs (at a fixed  M$_V$) is higher in systems that harbor intermediate age population. This is attributed to the contributions of the two channels in generating ACs. However, in purely old galaxies such as Sculptor or Eridanus~II, only the progeny of binary stars is expected.

The detection of CCs, ACs and RRLs in LG dwarf galaxies is of great importance not only to assess precise and accurate distances to those systems but also to be able to study the radial profiles of different populations --and therefore trace different ages--, measure their metallicity gradients and help reconstruct the star formation history in galaxies where using population synthesis methods (which requires reaching the main-sequence) is difficult or impossible.

In the following sections, I will highlight some of the results obtained from pulsating variable stars in classical, ultra-faint and ultra-diffuse dwarf galaxies. 
I further discuss how the identification of these pulsating stars provides valuable insights into the star formation history, distance determination and possible tidal disruption of the host galaxy. Finally, I will conclude this manuscript by providing some RRL period-luminosity relations in the SDSS pass-band system to show the potential of having an homogeneous distance calibrator in the near future with the advent of the Vera C. Rubin LSST survey.    

\section{Deciphering the star-formation with pulsating stars in classical dwarf galaxies}
\begin{figure*}[!ht]
\resizebox{\hsize}{!}{\includegraphics[clip=true]{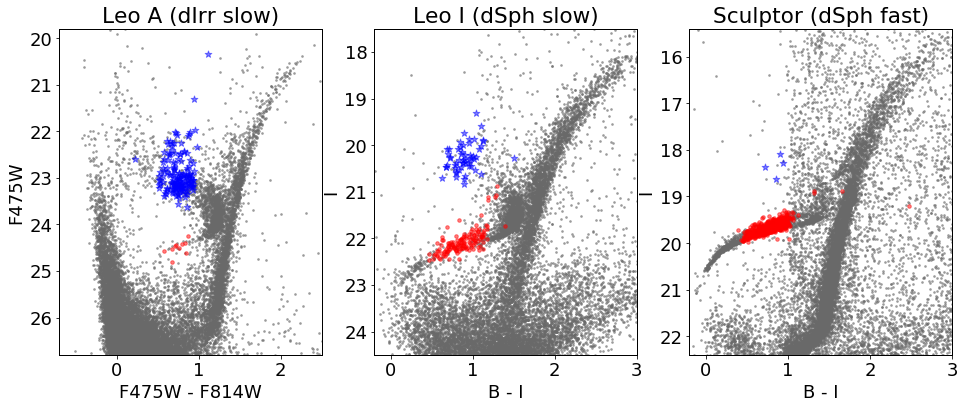}}
\caption{\footnotesize Color-magnitude diagrams of three classic dwarf galaxies (Leo~A, Leo~I and Sculptor) with different morphology and star formation histories. Highlighted are their RRLs (red dots) and Cepheids (blue stars), either CCs or ACs. From left to right, the dominant stellar component transitions from a notable young population to prominent old population. The ratio of RRLs over Cepheids increase also from left to right. It is evident how these pulsating stars prove invaluable in distinguishing among various systems that have experienced distinct star formation histories.}
\label{fig:cmd_dawrfs}
\end{figure*}
Dwarf galaxies are the most abundant type of galaxy in the Universe. The term \textit{dwarf galaxies} is used to refer to low luminosity galaxies (typically fainter that M$_V \sim –17$) dominated by dark matter (e.g., \citealt{Tolstoy2009}). The LG hosts a large number of dwarf galaxies of different morphological types: dwarf spheroidals (dSph; devoid of gas and with no recent star formation history, also called \textit{early types}), dwarf irregulars (dIrr; gas rich with an active star formation histories) and dwarf transition types (dT; with intermediate properties between the two groups). A classification based on the full star formation histories (SFHs) of dwarf galaxies defines two types of dwarf galaxies: slow and fast dwarf galaxies. \textit{Slow dwarfs} formed a small fraction of their stellar mass at an early epoch, and continued forming stars until the present \citep{Gallart2015}. All dIrr with available full SFHs can be classified as slow dwarfs. Also, some dSphs like Leo I, Fornax or Carina have important intermediate-age and even young populations, and thus SFHs that resemble those of dIrrs. These can also be classified as slow dwarfs. \textit{Fast dwarfs} are those that started their evolution with a dominant star formation event and their period of star formation activity was short (less than a few Gyr). Most dSph galaxies are fast dwarfs, but not all.

Numerous works have searched for variable stars within dwarf galaxies (see compilation of references for variable star studies in LG dwarf galaxies made by \citealt{MartinezVazquez2019} and \citealt{MonelliFiorentino2022}). In the context of this review, I will just mention three galaxies that combine different morphology classification and star-formation histories to show how different star-forming rates impact the population of CCs, ACs and RRLs.

Leo~A is a predominantly young dIrr galaxy with a considerable delay in its star formation history. \cite{Cole2007} refer to this galaxy as an extreme case of a ``late bloomer" in the LG since they find that over 90\% of all the star formation that ever occurred in Leo~A happened more recently than 8~Gyr ago (with the peak of the star formation rate happening at 1.5~Gyr) and ongoing star formation to this day \citep{Lescinskaite2022}.
\cite{Bernard2013} identified 156 Cepheids and only 10 RRLs. About the 90\% of the detected Cepheids have periods shorter than 1.5 d. In this case also, a comparison with theoretical models using evolutionary tracks for stars that ignite helium in the core in degenerate (ACs) and non-degenerate conditions (CCs) indicate that some of the fainter stars classified as CCs could be ACs. In comparison with Leo~I, the density of RRLs in Leo~A is very low. This agrees with the star-formation history obtained by \citet{Cole2007} where only a very small amount of star formation occurred in the first few Gyr after the Big Bang.

Leo~I is a dSph galaxy with a slow star-formation history. \cite{Fiorentino2012} study its pulsating stars and conclude that Leo~I contains the largest sample of Cepheids and the largest Cepheids to RRL ratio found in a dSph (106 RRL stars and 51 Cepheids). In addition, because of its extended and recent star formation history (star formation enhancements 13, 5.5, 2, and 1~Gyr ago, after which it was substantially quenched; \citealt{RuizLara2021}), its Cepheids traces a unique mix of ACs (blue extent of the red-clump, partially electron-degenerate central helium-burning stars) and short-period CCs (blue-loop, quiescent central helium-burning stars).

On the other hand, Sculptor is a dSph galaxy with a fast star-formation history. \cite{deBoer2012} showed that the star formation history of Sculptor is dominated by old ($>$10 Gyr) stars, and the majority of its total mass in stars was formed between 14 and 7~Gyr ago, with a peak at 13-14 Gyr ago. In terms of pulsating stars, the galaxy does not contain any CC (because of the absence of any recent star-formation), but according to the search for variable stars made by \cite{MartinezVazquez2016b}, it contains a plethora of RRL stars (536) and a few (4) ACs.

These three illustrative cases comprise two slow galaxies with different stars formation histories (an extended and a very recent one) and one fast dwarf galaxy. These cases help us realize how powerful detecting different types of pulsating stars --such as RRLs, ACs and CCs that belong to different stellar population-- are in order to determine the star formation history of the galaxies. Figure~\ref{fig:cmd_dawrfs} shows the color-magnitude diagrams for the three aforementioned galaxies, highlighting their RRLs and Cepheid stars.

\section{The role of pulsating stars in ``Newly discovered'' dwarf galaxies}
In the past two decades, we have witnessed the discovery of about 60 new ultra-faint, and two new ultra-diffuse, dwarf galaxies. These have been detected thanks to large-area, deep, multi-color imaging sky surveys carried out with the Dark Energy Camera (e.g., DES, SMASH, MagLites, DELVE), Hyper Suprime Cam, Pan-STARRS and Gaia. 
The low masses and stellar densities, coupled with the high contamination by field stars, make the determination of morphological parameters and distances for these galaxies a challenging task. A compelling approach to improve the distance determination to these ultra-faint systems --and thus clarify their nature-- is to detect RRL members. Moreover, RRLs can also provide valuable insight into the properties of the old stellar population of the host.

\subsection{Ultra-faint dwarf galaxies}
The Sloan Digital Sky Survey (SDSS) discovered a new class of objects, the \textit{ultra-faint} dwarf (UFD) galaxies, the first examples being Willman~1 and Ursa Major~I \citep{Willman2005a,Willman2005b}. These UFDs extend the spectrum of properties of \textit{classical} LG dwarf galaxies to a lower mass regime (M$_V > -7.7$ mag, Simon 2019). Since these first discoveries, more than 50 UFDs have been found in the Milky Way (MW) neighbourhood \citep[see e.g.][and references therein]{DrlicaWagner2020,Cerny2023b}. UFDs are possibly the oldest and most primitive of galaxies \citep{Bose2018,Simon2019}. According to the hierarchical galaxy formation model \citep{WhiteFrenk1991}, large galaxies are built up by the accretion of smaller galaxies; thus, UFDs may be representative of the basic building blocks of the galaxy formation process. 

There are several studies in the literature searching for variable stars in UFDs. UFDs are prone to harbor at least one RRL if the absolute $V$ magnitude of the host system is M$_V \lesssim -3.5$ \citep{MartinezVazquez2019}. Table~\ref{tab:rrl_ufd_updated} shows the updated number of RRL stars (up to the year 2023) associated to ultra-faint systems sorted by their heliocentric distances.
\begin{table*}[!ht]
\caption{Compilation of the number of RRL stars found in ultra-faint dwarf (UFD) systems up to the year 2023.}
\begin{scriptsize}
\begin{tabular}{lrrrl}
\hline
\textbf{Name} & \textbf{$\alpha_{J2000}$} (deg) & \textbf{$\delta_{J2000}$} (deg) & \textbf{$\rm{N_{RRL}}$} & \textbf{References} \\
\hline
Dra2   & 238.198333 & 64.565278  &  0 & \citet{Vivas2020b}\\ 
Tuc3   &  359.1075 & -59.58332   &  6 & \citet{Vivas2020b}\\ 
Seg1   &  151.7504 &  16.0756    &  1 & \citet{Simon2011}\\ 
Hyi1   &   37.3890 & -79.3089    &  4 & \citet{Koposov2018, Vivas2020b}\\ 
Car3   &  114.6298 & -57.8997    &  0 & \citet{Torrealba2018}\\ 
Tri2   &   33.3252 &  36.1702    &  0 & \citet{Vivas2020b}\\ 
Cet2   &   19.4700 & -17.4200    &  0 & \citet{Vivas2020b}\\ 
Ret2   &   53.9203 & -54.0513    &  0 & \citet{Vivas2020b}\\ 
UMa2   &  132.8726 &  63.1335    &  4 & \citet{DallOra2012, Vivas2020b}\\ 
Seg2   &   34.8226 &  20.1624    &  1 & \citet{Boettcher2013}\\ 
Car2   &  114.1066 & -57.9991    &  3 & \citet{Torrealba2018}\\ 
Will1  &  162.3436 &  51.0501    &  0 & \citet{Siegel2008}\\ 
Boo2   &  209.5141 &  12.8553    &  1 & \citet{Sesar2014}\\ 
CBer   &  186.7454 &  23.9069    &  3 & \citet{Musella2009, Vivas2020b}\\ 
Boo3   &  209.3000 &  26.8000    &  7 & \citet{Sesar2014, Vivas2020b}\\ 
Tuc4   &  0.717000 & -60.8300    &  1 & \citet{Stringer2021}\\ 
Tuc5   &  354.3470 & -63.2660    &  0 & \citet{Vivas2020b}\\ 
Gru2   &  331.0250 & -46.4420    &  1 & \citet{MartinezVazquez2019}\\ 
Tuc2   &  342.9796 & -58.5689    &  3 & \citet{Vivas2020b}\\ 
Boo1   &  210.0200 &  14.5135    & 16 & \citet{DallOra2006,Siegel2006, Vivas2020b}\\ 
Sag2$^{*}$  & 298.168750 & -22.068056 &  6 & \citet{Joo2019, Vivas2020b}\\ 
DELVE2 &   28.7720 & -68.2530    &  0 & \citet{Cerny2021a}\\ 
Eri4   &   76.4380 &  -9.5150    &  1 & \citet{Cerny2021b}\\ 
Hor2   &   49.1077 & -50.0486    &  0 & \citet{Vivas2020b}\\ 
Hor1   &   43.8813 & -54.1160    &  0 & \citet{Vivas2020b}\\ 
Phe2   &  354.9960 & -54.4115    &  1 & \citet{MartinezVazquez2019}\\ 
Peg4   &  328.5390 &  26.6200    &  3 & \citet{Cerny2023a}\\ 
Vir1   &  180.0380 &  -0.6810    &  0 & \citet{Vivas2020b}\\ 
Eri3   &  35.695200 & -52.2838   &  1 & \citet{Vivas2020b}\\ 
Ret3   &   56.3600 & -60.4500    &  1 & \citet{Vivas2020b}\\ 
UMa1   &  158.7706 &  51.9479    &  7 & \citet{Garofalo2013}\\ 
Kim2$^{*}$  & 317.2020 & -51.1671     &  0 & \citet{MartinezVazquez2019}\\ 
Aqu2   &  338.4813 &  -9.3274    &  1 & \citet{Hernitschek2019}\\ 
Cen1   &  189.5850 & -40.9020    &  3 & \citet{MartinezVazquez2021c}\\ 
Gru1   &  344.1660 & -50.1680    &  2 & \citet{MartinezVazquez2019}\\ 
Her    &  247.7722 &  12.7852    & 12 & \citet{Musella2012, Garling2018}\\ 
Hyd2   &  185.4251 & -31.9860    &  1 & \citet{Vivas2016}\\ 
Leo4   &  173.2405 &  -0.5453    &  3 & \citet{Moretti2009}\\ 
CVen2  &  194.2927 &  34.3226    &  2 & \citet{Greco2008}\\ 
Leo5   &  172.7857 &   2.2194    &  3 & \citet{Medina2017}\\ 
Pis2   &  344.636458 & 5.955544  &  2 & \citet{Garofalo2021}\\ 
Peg3   &  336.107417 & 5.415047  &  2 & \citet{Garofalo2021}\\ 
Cet3   &   31.3310 &  -4.2700    &  1 & \citet{Stringer2021}\\ 
Eri2   &   56.0925 & -43.5329    & 67 & \citet{MartinezVazquez2021b}\\ 
And16  &   14.87625 & 32.37611   &  8 & \citet{MartinezVazquez2017, Monelli2016}\\ 
And13  &   12.96250 & 33.00444   &  9 & \citet{Yang2012}\\ 
And11  &   11.582080 & 33.80278  & 15 & \citet{Yang2012}\\ 
\hline
\end{tabular}
\begin{itemize}
\item[] \textit{Notes}.- The table is sorted by heliocentric distances, ranging from 21.5 kpc (Dra2) to 763 kpc (And11). Systems marked with an asterisk ($^{*}$) are now confirmed clusters.
\end{itemize}
\label{tab:rrl_ufd_updated}
\end{scriptsize}
\end{table*}
The motivation to search for RRL stars in these ultra-faint systems is to get precise distance measurements to help improve estimates of parameters such as the absolute visual magnitude and the physical size of the host. As an example, the first detection of RRL stars in Grus~I (2), Phoenix~II (1), and in Grus~II (1) allowed \cite{MartinezVazquez2019} to measure distances of these galaxies for the first time using stellar candles, which placed them farther away than predicted in their discovery papers. These refined distance implied larger sizes for these systems with a 30\% change for Phoenix II and 5\% for Grus~I and Grus~II. 
\begin{figure*}[!ht]
\resizebox{\hsize}{!}{\includegraphics[clip=true]{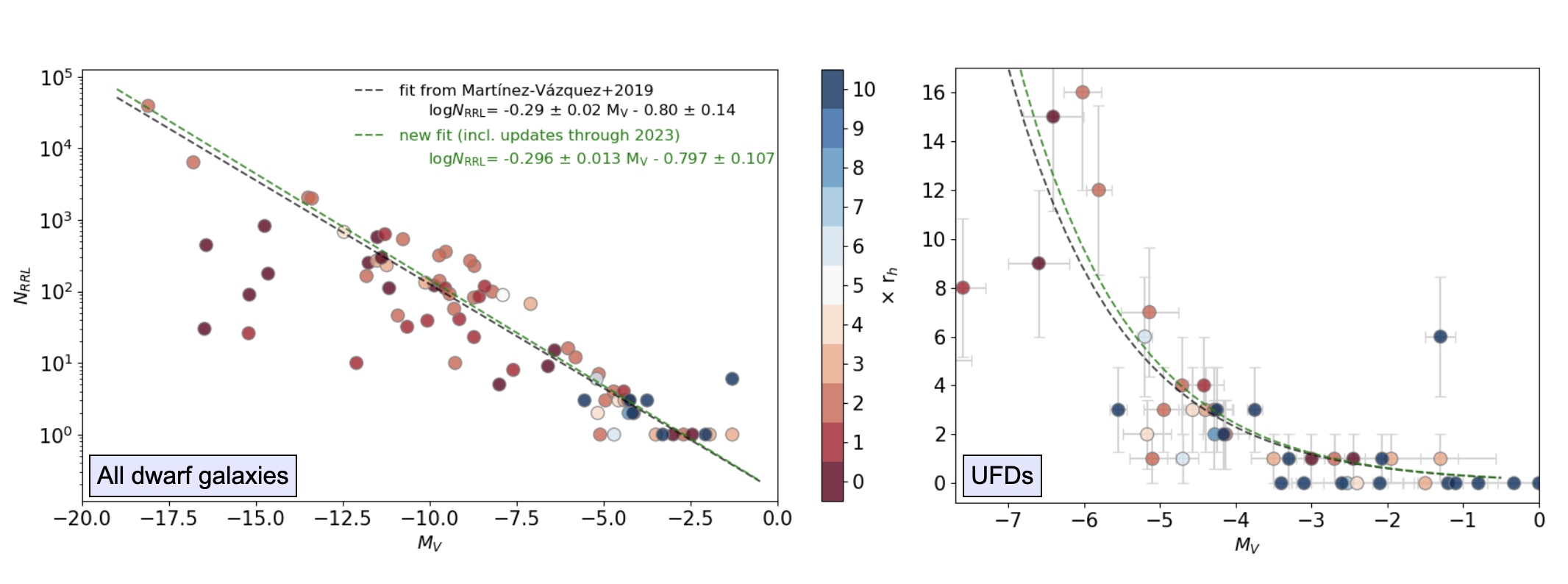}}
\caption{\footnotesize Number of RRL stars in dwarf galaxies as a function of the absolute $V$ magnitude of the host galaxy, $\rm{M_V}$. The different colors represent the area (in $\rm{r_h}$ units) of the variable star search for a given galaxy. The black and green dashed lines show the linear fit between $\log{N_{RRL}}$ and $\rm{M_V}$ for those galaxies where the variability search areas was greater or equal than 2\,$\rm{r_h}$ obtained by \citet{MartinezVazquez2019} and this work after including updates (up to the year 2023). The left panel is in semi-logarithmic scale while the right panel is a zoom-in of the left panel on the faint dwarf regime ($\rm{M_V} > -7.7$ mag) in a linear scale. The black and green lines represent the same respective fits in the left and right panels.}
\label{fig:rrl_relation}
\end{figure*}
RRL stars can also be instrumental in studying the extent of the systems and identifying potential tidal stripping. \citet{Garling2018} identified three extra-tidal RRL stars in Hercules in addition to its nine already known RRL stars within its tidal radius \citep{Musella2012}, which suggest that Hercules has been tidal stripped. In particular, the location of one of the outer RRLs is aligned with Hercules' orbit and consistent with the debris being in that direction. 

The most exhaustive investigation concerning the association of RRL stars with UFDs was carried out by \cite{Vivas2020b} using the Gaia DR2 RRL catalog \citep{Clementini2019}. \cite{Vivas2020b} found 47 RRL stars associated with 14 different MW ultra-faint satellites within 100 kpc. They identified RRLs for the first time in Tucana~II, finding additional members in Ursa Major~II, Coma Berenices, Hydrus~I, Bootes~I and Bootes~III, and distinguishing possible candidate extra-tidal RRLs in Bootes~I, Bootes~III, Sagittarius~II (cluster), Tucana~III, Eridanus~III (unclassified system), and Reticulum~III. Recently, several independent studies kept searching and detecting RRL stars as member of newly discovered galaxies (DELVE~2: \citealt{Cerny2021a}, Centaurus~I: \citealt{MartinezVazquez2021c}, Eridanus~IV: \citealt{Cerny2021b}, Pegasus~IV: \citealt{Cerny2023a}) or already know UFDs (Pisces~II, Pegasus~III: \citealt{Garofalo2021}, Eridanus~II: \citealt{MartinezVazquez2021b}), Cetus~III, Tucana~IV: \citealt{Stringer2021}). In these studies, distant RRL stars have been identified at ${\sim\,}6\,\rm{r_h}$ from Tucana~IV and Centaurus~I and ${\sim\,}10\,\rm{r_h}$ from Pegasus~IV. The detection of the aforementioned distant RRLs may suggests that a past tidal disruption could have happened in these galaxies. Nonetheless, to confirm this scenario, radial velocities of these stars are needed to determine whether those RRLs are (or are not) members of those UFDs.

Figure~\ref{fig:rrl_relation} shows a comprehensive analysis up to the year 2023 of the number of RRLs in dwarf galaxies. This figure is an updated version of Figure 10 in \cite{MartinezVazquez2019} and it includes as a color-map the information about the variable star search area (in $\rm{r_h}$ units) for each galaxy. An updated fit of the number of RRL stars as a function of $\rm{M_V}$ for those galaxies where the variability search covers an area greater or equal than 2$\rm{r_h}$ (green dashed line in Figure~\ref{fig:rrl_relation}) is given by:
\begin{equation}\label{eq:NRRL_MV}
\log{N_{\rm{RRL}}}= - 0.296 \pm 0.013 M_{\rm{V}} - 0.797 \pm 0.107    
\end{equation}
which is nearly identical to the relation provided by \citet{MartinezVazquez2019} (black dashed line in Figure~\ref{fig:rrl_relation}) but with slightly improved precision in the coefficients.

Either by utilizing this correlation or by examining Figure~\ref{fig:rrl_relation}, we can deduce that the discovery of new UFDs through the identification of cluster of RRLs \citep{BakerWillman2015} will be effective only for those UFDs brighter than $\rm{M_V}{\,\sim\,}{-}4$ mag. \citet{Stringer2021} quantified the sensitivity of this search using a suite of simulated satellite galaxies generated by \citet{DrlicaWagner2020} where the expected number of RRLs was predicted by using the previous version of Eq.~(\ref{eq:NRRL_MV}). They found that an RRL-based search is more sensitive than those using resolved stellar populations in the regime of large ($\rm{r} >$ 500~pc) and low surface-brightness dwarf galaxies. However, the isochrone matched-filter searches remain more sensitive for satellites with $\rm{M_V}{\,>\,}-5$~mag, due to the small number of RRLs expected in these galaxies.
\begin{figure*}[t!]
\resizebox{\hsize}{!}{\includegraphics[clip=true]{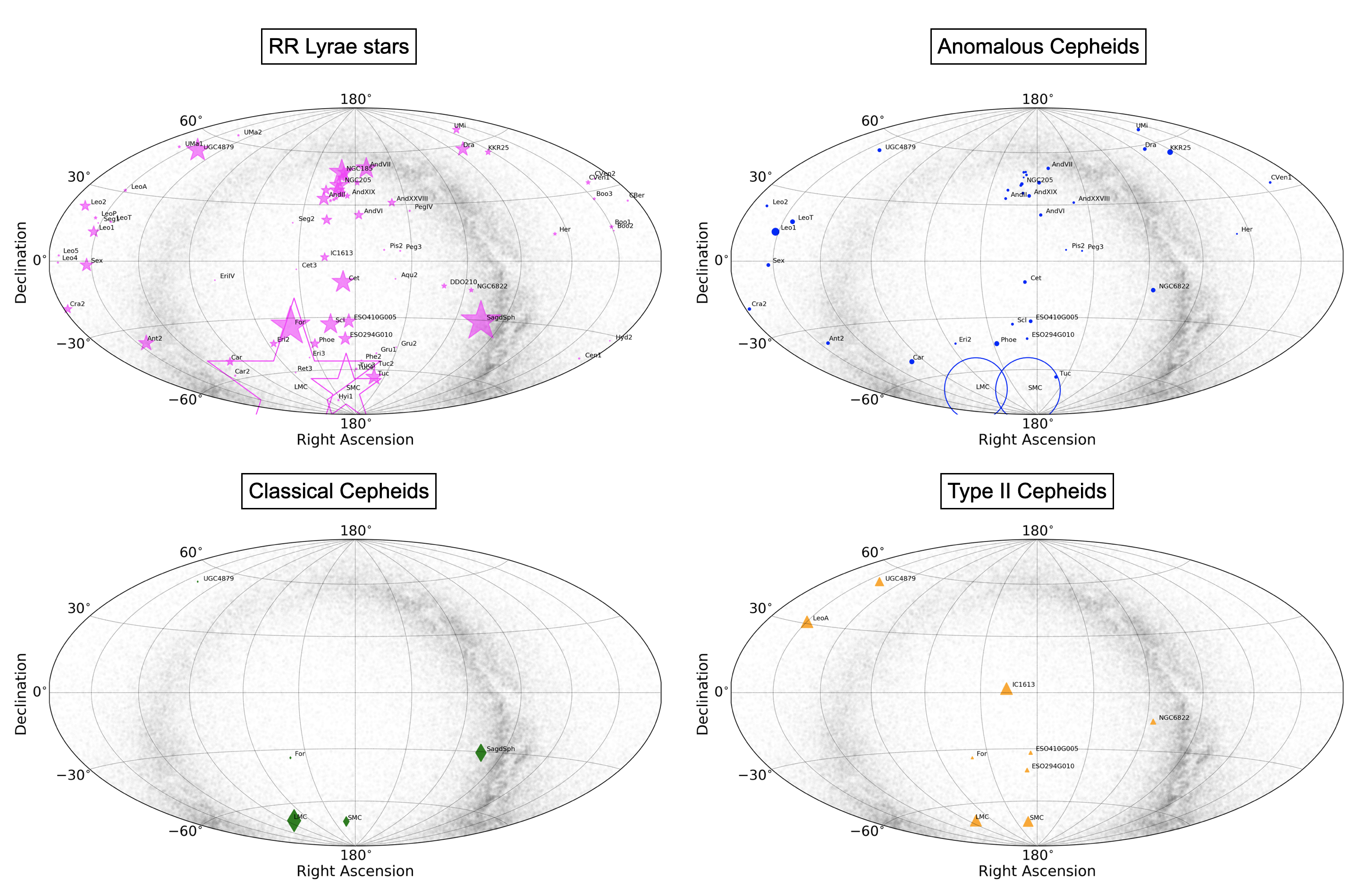}}
\caption{\footnotesize Census of RR Lyrae stars (upper left), anomalous Cepheids (upper right), classical Cepheids (bottom left) and type II Cepheids (bottom right) in LG dwarf galaxies. The size of the symbols is a representation of the number of stars of each type that a galaxy contains. The upper panels depict the LMC and SMC as unfilled symbols to enhance visualization. As both galaxies are known for harboring a substantial population of RR Lyrae stars and anomalous Cepheids, their symbols overlap with others, and if filled, they would obscure them.}
\label{fig:census}
\end{figure*}
\begin{figure*}[t!]
\hspace{-2cm}
\resizebox{1.3\hsize}{!}{\includegraphics[clip=true]{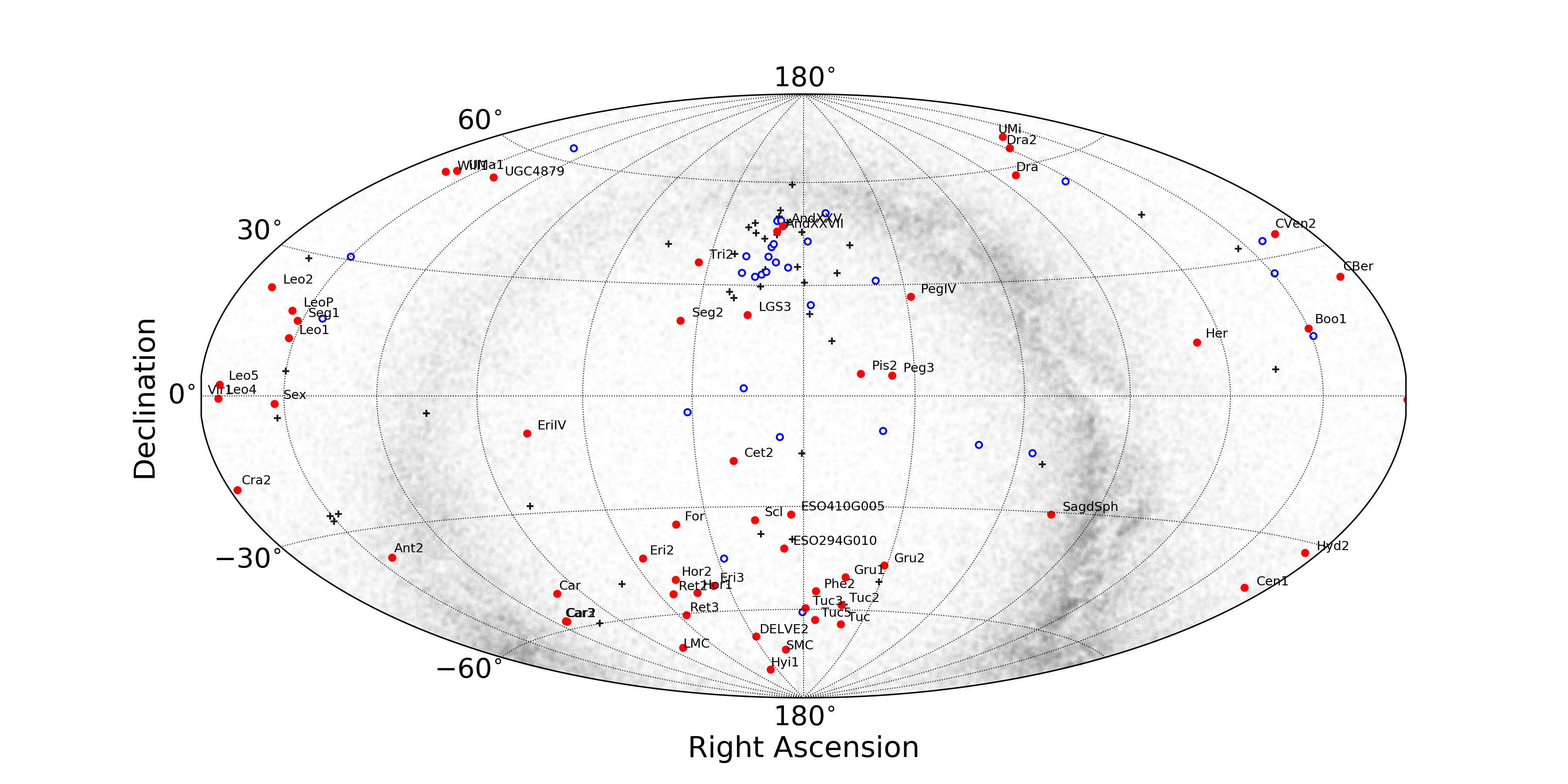}}
\caption{\footnotesize Distribution of dwarf galaxies in the sky up to $\sim$2~Mpc. The red dot symbols are those galaxies where variable star searches have been conducted beyond 2\,$\rm{r_h}$. Blue circles are those that do not meet the above criteria but where variable searches have been carried out. Black crosses are galaxies with no variability studies to the date. To allow the readability of the figure, only galaxies with variable star searches covering an area larger than 2\,$\rm{r_h}$ have been labeled.
}
\label{fig:map}
\end{figure*}

\subsection{Ultra-diffuse dwarf galaxies}
\citet{Torrealba2016, Torrealba2019} discovered two new MW dwarf galaxies (Crater~II and Antlia~II) of a type unprecedented in the MW vicinity, the so-called \textit{ultra-diffuse galaxies} (UDG). These galaxies have stellar masses comparable to classical dwarfs like Sculptor or Fornax but they are much more extended systems. Their surface brightnesses are fainter than 30~mag~arcsec$^{-2}$. Recent publications show that these galaxies host hundreds of RRL stars (99 Crater~II, \citealt{Joo2018,Vivas2020a}; 318 Antlia~II; \citealt{Vivas2022}) plus several ACs (7 Crater~II, 8 Antlia~II) located within 2~r$_h$.

From the spatial distribution of the RRLs in Crater~II, \citet{Vivas2020a} detect a more elongated shape ($\epsilon = 0.24$) than found when selecting giant branch stars ($\epsilon = 0.12$). From their high quality light curves, they refine the distance of Crater II at 117 $\pm$ 4 kpc and measure a metallicity dispersion of 0.2 dex, consistent with that found for the RGB members stars by spectroscopy \citep{Caldwell2017,Fu2019}, and with its relatively narrow red giant branch \citep{Walker2019}.

Antlia~II also shows an elongated shape ($\epsilon = 0.28$) from its population of RRLs \citep{Vivas2022}, although a more extended shape has been measure by their spectroscopic members ($\epsilon = 0.60$, \citealt{Ji2021}). It is slightly farther (average distance to Antlia~II based on the RRLs is 124.1~kpc), more massive and extended than Crater~II ($\rm{r_h}$ is $\sim 3$ times larger), and it exhibits a distance gradient of 2.72 kpc deg$^{-1}$ along the semi-major axis of the galaxy -- with the southeast side being farther away than the northwest side. The elongation along the line of sight is likely due to the ongoing tidal disruption of Antlia~II. In particular, \citet{Vivas2022} show that a model in which Antlia~II is tidally disrupting explains the observed distance gradient from the RRLs.

\section{Updated census of pulsating stars}
Using the compilation from \citet{MartinezVazquez2019, Vivas2020b} and recent updates (see Table~\ref{tab:rrl_ufd_updated}) for RRL stars and the compilation of \citet{MonelliFiorentino2022} for CCs, ACs and type II Cepheids, Figure~\ref{fig:census} shows the amount of each of these pulsating stars (weighted by the symbol size) in LG dwarf galaxies. Looking at these maps, one can clearly notice how RRLs are the most abundant pulsating stars detected in the LG.

Figure~\ref{fig:map} displays the distribution of all the known dwarf galaxies up to $\sim$2~Mpc. The chart highlights whether variability studies have been carried out in these galaxies or not. To date, there are 129 dwarf galaxies within 2~Mpc and to the best of my knowledge variability studies have been carried out in 88 of them. However, there are only 57 dwarf galaxies where their variable star searches span an area up to or beyond 2 half-light radii. From the 41 without variable studies yet, only 5 of them have  distance moduli below 20.5~mag and thus can be searched with Gaia for RRL stars.

\section{Towards an homogeneous RR Lyrae distance scale}
In the past few decades, the use of the SDSS pass-bands has become significant in big survey \textit{cannons} such as PanSTARRS and DECam. The upcoming Vera Rubin LSST survey will also observe in the $ugrizy$ bands.

Studies based on star formation histories of LG galaxies show that all systems contain an old population \citep[see e.g.,][]{Weisz2014}. Since old populations are ubiquitous, RRL stars can be found almost everywhere in our Local neighborhood. 

In order to get precise and homogeneous distances using RRL stars, it is crucial to build period-luminosity relations in the SDSS bands. Recently, \cite{Marconi2022} provided new theoretical period-luminosity-metallicity relations for the Vera C. Rubin LSST filters. 

However, only a few empirical period-luminosity relations in the SDSS bands have been derived \citep{Sesar2017,Vivas2019,MartinezVazquez2021b}, utilizing systems that exhibit distinct metallicity distributions.

\begin{figure}[t!]
\hspace{-0.45cm}
\resizebox{1.1\hsize}{!}{\includegraphics[clip=true]{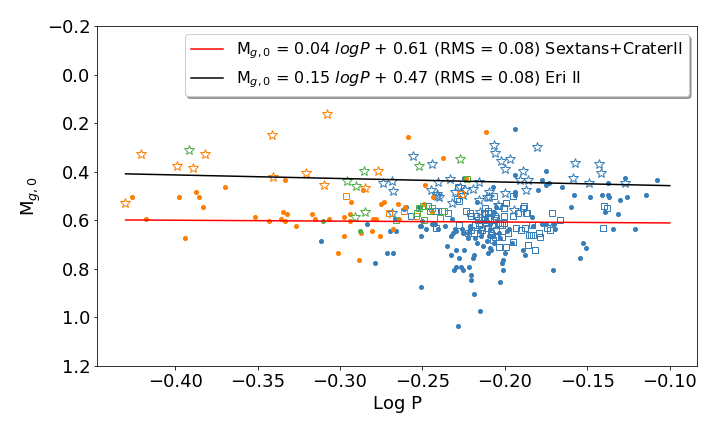}}

\hspace{-0.45cm}
\resizebox{1.1\hsize}{!}{\includegraphics[clip=true]{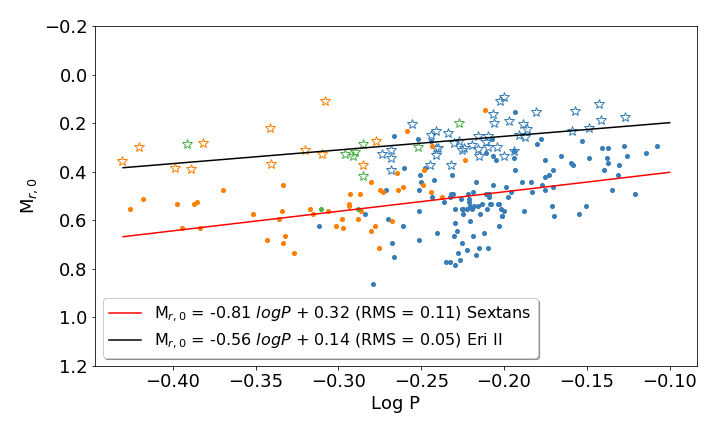}}

\hspace{-0.45cm}
\resizebox{1.1\hsize}{!}{\includegraphics[clip=true]{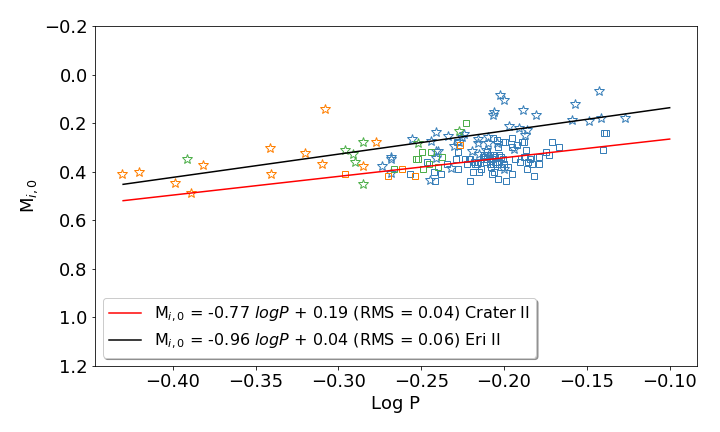}}

\caption{\footnotesize Period versus absolute magnitude plots in the $g,r,i$ bands for different dwarf galaxies observed with DECam (dots: Sextans, squares: Crater~II, stars: Eridanus~II). The different colors represent the different types of RRLs (blue: RRab, orange: RRc, green: RRd). The periods of the RRcd stars have been fundamentalized. The solid lines are the empirical period-luminosity relations derived by fitting the data (black: Sextan, Crater~II or both ($\langle \rm{[Fe/H]} \rangle \approx -2.0$); red: Eridanus~II ($\langle \rm{[Fe/H]} \rangle \approx -2.4$).}
\label{fig:plr}
\end{figure}

Figure~\ref{fig:plr} presents the period-luminosity relations of the only three galaxies with enough RRL stars and well-sampled light curves observed with DECam in the $g,r,i$ bands (Sextans: \citealt{Vivas2019}, Crater~II: \citealt{Vivas2020a}, and Eridanus~II: \citealt{MartinezVazquez2021b}). The zero-point differences are mainly driven by the metallicity dependence. Since Sextans and Crater~II share the same mean metallicity of [Fe/H] $\sim -2.0$~dex \citep{Battaglia2011,Ji2021} both data set overlap and share a common period-luminosity relation in the g-band (upper panel). Eridanus~II is a more metal-poor system, [Fe/H] $\sim -2.4$~dex \citep{Li2017}. This is reflected in its fainter absolute magnitudes, i.e., larger zero-point.

Deriving a well-calibrated set of period-luminosity relations in the SDDS bands that spans various metallicity and period ranges is essential as the Vera C. Rubin LSST survey begins detecting RRL stars, aiming for accurate distance measurements. This effort will enhance our ability to gain a more detailed understanding of the Galactic structure.

\section{Conclusion and Final Remarks}
About 130 known dwarf galaxies within a heliocentric distance of 2~Mpc reside in our LG. Out of them 44\% have a catalog of variable stars that at least reach 2\,r$_h$ and 32\% have not yet been searched for variables (see Figure~\ref{fig:map}).

We are continuing our observations on these interesting and important systems, and anticipate a harvest of new galaxies to study following the commencement of the Vera C. Rubin LSST survey.

Additionally, future imaging surveys will lead to the detection of new UFDs and UDGs \citep[see, e.g.,][]{Mutlu-Pakdil2021}. The characterization of these systems is of great value to study the dark matter \citep{Nadler2021} and trace the mass assembly history of the MW. The determination of accurate and precise distances is crucial in determining the nature (mass and size), orbits, and pericentric passages of these Galactic building blocks.

The advent of extremely large telescopes allow the detection of RRL stars up to 6~Mpc and, therefore, will enable the study of old populations in systems where the main-sequence cannot be resolved. As shown in this review, the identification and analysis of pulsating stars that belong to different stellar populations such as Cepheids and RRLs, will be crucial to decipher the star-formation history and to investigate age gradients in their host galaxies.

\begin{acknowledgements}
C.E.M.-V. thanks all her collaborators and many other researchers for their contribution and hard work searching for pulsating variable stars in LG galaxies. C.E.M.-V. also expresses gratitude to Andrew Pace for sharing his updated compilation of Local Volume dwarf galaxies properties, John Blakeslee for his careful reading of the manuscript and valuable comments, and the referee for reviewing the manuscript and  providing valuable insights.\\ 

C.E.M.-V. is supported by the international Gemini Observatory, a program of NSF's NOIRLab, which is managed by the Association of Universities for Research in Astronomy (AURA) under a cooperative agreement with the National Science Foundation, on behalf of the Gemini partnership of Argentina, Brazil, Canada, Chile, the Republic of Korea, and the United States of America.\\
\end{acknowledgements}

\bibliographystyle{aa}

\end{document}